\documentclass{Interspeech}

\usepackage[flushleft]{threeparttable}
\usepackage{booktabs,ragged2e}
\usepackage{booktabs}
\usepackage{bm}
\usepackage{multirow}
\usepackage{hyperref}
\usepackage{pifont}
\usepackage{subfig}

\interspeechcameraready

\title{Speech transformer models for extracting information from baby cries}

\author[affiliation={1,2}]{Guillem}{Bonafos}
\author[affiliation={1}]{Jérémy}{Rouch}
\author[affiliation={1}]{Lény}{Lego}
\author[affiliation={1,5}]{David}{Reby}
\author[affiliation={3}]{Hugues}{Patural}
\author[affiliation={1,4,5}]{Nicolas}{Mathevon}
\author[affiliation={2,5}]{Rémy}{Emonet}

\affiliation{ENES Bioacoustics Research Lab}{$^2$Laboratoire Hubert Curien, $^3$SAINBIOSE Laboratory, $^4$CHArt Lab}{$^5$Institut Universitaire de France}

\email{first.last@univ-st-etienne.fr}

\usepackage{comment}

\begin{document}

\maketitle

\begin{abstract}
    
    Transfer learning using latent representations from pre-trained speech models achieves outstanding performance in tasks where labeled data is scarce. However, their applicability to non-speech data and the specific acoustic properties encoded in these representations remain largely unexplored. In this study, we investigate both aspects. We evaluate five pre-trained speech models on eight baby cries datasets, encompassing 115 hours of audio from 960 babies. For each dataset, we assess the latent representations of each model across all available classification tasks. Our results demonstrate that the latent representations of these models can effectively classify human baby cries and encode key information related to vocal source instability and identity of the crying baby. In addition, a comparison of the architectures and training strategies of these models offers valuable insights for the design of future models tailored to similar tasks, such as emotion detection.
\end{abstract}

\section{Introduction}

A baby's cry is a signal that conveys a wealth of information such as the child's identity \cite{gustafsson_fathers_2013, bouchet_baby_2020}, age \cite{lockhart-bouron_infant_2023} and the intensity of discomfort or pain \cite{koutseff_acoustic_2018}. Furthermore, it holds potential as a biomarker for detecting disorders and dysfunctions \cite{orlandi_automatic_2012, esposito_cry_2017, khozaei_early_2020, lawford_acoustic_2022}. Developing a representation of cries that maximizes the extraction of such information is therefore of paramount importance.

Studies on cries have mainly focused on the computation of acoustic features. However, transfer learning from self-supervised pre-trained models has emerged as the state of the art in many speech-related tasks \cite{mohamed_self-supervised_2022}. These models, which leverage attention mechanisms \cite{vaswani_attention_2017}, are generally trained through self-supervision on large speech datasets. They enable the learning of high-quality representations, even in scenarios with limited data availability, and have proven effective for analyzing vocalizations from other species \cite{cauzinille_investigating_2024}. 
Extending the use of these representations to the analysis of cries would be particularly valuable, given the high costs associated with collecting labeled data in this domain.

The study of baby cries using these pre-trained models may offer a second advantage by providing insight into what is encoded within the representations learned by these speech models. For instance, the pain expressed in a cry is embedded in non-linear acoustic phenomena (roughness) \cite{cornec_human_2024, corvin_pain_2024}. Thus, assessing the ability of these representations to determine the discomfort or pain level of a cry may serve as a means to evaluate their ability to encode these parameters, which reflect the instability of the vocal source \cite{anikin_nlp}.

We hypothesize that the representations of pre-trained speech models can be used to analyze infant cries. Specifically, we propose that these representations encode information about the non-linear phenomena of vocalizations. To test this hypothesis, we compute latent representations from five pre-trained speech models on eight cry datasets from different previous studies. Then, we apply a probing methodology to evaluate each task and each representation. Performance is compared across models for each classification task, as well as differences between the various models.

\section{Methods}\label{part:methods}

\subsection{Datasets and classification tasks}\label{part:datasets}


To assess the ability of latent representations from pre-trained speech models to classify baby cries, we consolidated eight available datasets. Each dataset comprises several cry recordings, labeled differently according to the respective study. These labels defined our probing tasks. Table~\ref{tab:datasets} provides an overview of the analyzed datasets and the distribution of audio durations. It should be noted that the datasets do not contain uniform information, resulting in different classification tasks from one set to another. The centralization of these datasets, along with the publications from which they originate, allows us to establish a benchmark for the expected results.

\begin{table}
\fontsize{8pt}{8pt}\selectfont
\caption{Datasets and their original publication. Two of them have not been published yet. Donate A Cry can be found on Github. We give the total number of cry, their mean duration and standard deviation, in seconds, the number of baby recorded and the range of cries per baby. For Donate A Cry dataset, we do not have the info about the identity of the baby.}
\label{tab:datasets}
\setlength\tabcolsep{0pt} 

\newcommand{\PM}{$&$\pm$&$}

\begin{tabular*}{\linewidth}{@{\extracolsep{\fill}} c crclcc}
\toprule
     Dataset & $N_\text{cries}$ & \multicolumn{3}{c}{Duration (seconds)} & $N_\text{baby}$ & Range (per baby) \\ 
\midrule
     EnesBabyCries2\cite{lockhart-bouron_infant_2023} & 2024 & $51.33$ &$\scriptstyle\pm$& $85.22$ & 23 & \numrange{32}{261} \\
     Koutseff\cite{koutseff_acoustic_2018} & 321 & $7.29 \PM 1.15$ & 25 & \numrange{7}{25} \\
     Cornec\cite{cornec_human_2024} & 72 & $2.17 \PM 1.52$ & 36 & $2$ \\
     Bouchet\cite{bouchet_baby_2020} & 248 & $7.42 \PM 2.27$ & 31 & \numrange{1}{19}  \\
     Lefkir & 719 & $13.42 \PM 11.74$ & 41 & \numrange{3}{36} \\  
     Vial & 794 & $104.70 \PM 86.73$ & 18 & \numrange{5}{221} \\
     Donate A Cry & 457 & $6.92 \PM 0.11$ & Unknown & Unknown \\
     CryCeleb\cite{budaghyan_cryceleb_2024} & 26093 & $0.90 \PM 0.56$ & 786 & \numrange{1}{143} \\
\bottomrule
\end{tabular*}

\end{table}

The EnesBabyCries2 dataset, from \cite{lockhart-bouron_infant_2023} and freely accessible\footnote{\url{https://osf.io/ru7na/?view_only=}}, contains longitudinal recordings of cries from 24 children. It includes information about the child's age at the time of recording (15 days, 1.5 months, 2.5 months or 3.5 months), the sex, identity and the cause (discomfort, hunger or isolation).
The Bouchet, Lefkir, and Vial datasets provide information on identity and sex. Of these, only the Bouchet dataset has been published \cite{bouchet_baby_2020}, while the other two have yet to be released. The Koutseff \cite{koutseff_acoustic_2018} and Cornec \cite{cornec_human_2024} datasets include additional information, if the cry is a cry of pain (cry during vaccine) or not (cry during bath), plus the type of vaccine for the Koutseff dataset. All these datasets will be made freely accessible in an upcoming publication, currently in preparation.
The Donate A Cry dataset\footnote{\url{https://github.com/gveres/donateacry-corpus/}} includes information on the cause of cry, age, and sex. CryCeleb \cite{budaghyan_cryceleb_2024} provides labels associated with the child's identity and age (birth or discharge).

Given the results of these previous studies, we expect no significant information related to sexes or causes. If the latent representations preserve relevant information, we should be able to predict whether the cry is associated with pain (characterized by non-linear phenomena, particularly roughness \cite{koutseff_acoustic_2018, cornec_human_2024, corvin_pain_2024}) and the identity and age of the crying individual (not associated with a particular acoustic parameter but by a combination of them \cite{lockhart-bouron_infant_2023}). If this proves to be the case, it would suggest that the latent representations encode the stability of the vocal source.

\subsection{Speech transformer models}\label{part:models}

To evaluate the performance of pre-trained speech models, we selected five models with varying architectures and training datasets, summarized in Table~\ref{tab:models}. Given the differences in architectures and training data, comparing their results can provide valuable insights into the optimal modeling choices for similar tasks.

\begin{table}
\fontsize{8pt}{8pt}\selectfont
\caption{Description of the models. All the models are transformers. The number of parameters is in million. We give the dimension of the latent representation that we use for the classification tasks.}
\label{tab:models}
\setlength\tabcolsep{0pt} 

\begin{tabular*}{\linewidth}{@{\extracolsep{\fill}} c cccc}
\toprule
     Model & $N_\text{parameters}$ & $D_\text{representation}$ & Training task & Input \\ 
\midrule
    Wav2Vec2 \cite{baevski_wav2vec_2020} & 317m & $(t, 1024)$ & contrastive & Wave \\
    HuBERT \cite{hsu_hubert_2021} & 317m & $(t, 1024)$ & predictive & Wave \\
    Whisper \cite{radford_robust_2023} & 1550m & $(1500, 1024)$ & supervised & Log-Mel \\
    Unispeech \cite{wang_unispeech_2021} & 317m & $(t, 1024)$ & contrastive+supervised & Wave \\
    AST \cite{gong_ast_2021} & 307m & $(1024, 768)$ & supervised & Log-Mel\\
\bottomrule
\end{tabular*}

\end{table} 

Wav2Vec2 \cite{baevski_wav2vec_2020} and HuBERT \cite{hsu_hubert_2021} are both pre-trained using self-supervised learning. Wav2Vec2 relies on a contrastive task trained on 53k hours of Librivox data, while HuBERT employs a predictive task trained on 60k hours of Libri-light data. Comparing their results will allow us to evaluate the capacity of self-supervised speech representations to extract meaningful information from infant cries.
Whisper \cite{radford_robust_2023}, on the other hand, is pre-trained on speech using multi-task supervised learning. Its training data is significantly larger, albeit potentially noisier, with over 5 million weakly labeled hours. Whisper's goal is to provide representations that can be used across multiple languages without fine-tuning, allowing us to assess the transferability of supervised learning to non-speech audio tasks.
Unispeech \cite{wang_unispeech_2021} builds upon the Wav2Vec2 architecture for self-supervised representation learning but incorporates a supervised task, enabling explicit alignment of the representations to known phonetic units. Its pre-training leverages 5k hours of CommonVoice data.
Finally, AST \cite{gong_ast_2021} is trained in a supervised manner using datasets such as AudioSet, ESC50, and Speech Commands. Unlike the other models, it does not use CNNs; instead, its initial layers are transferred from a pre-trained ImageNet model. While its training data is still human sounds, AST is the model most distantly related to speech among those compared in this study. Models are not fine-tuned.
Finally, we average each embedding over the time dimension to construct an embedding from each model.

\subsection{Probing methodology}

Probing is designed to assess the information encoded in the latent representations of the models described in Section~\ref{part:models}. 
The eight datasets presented in Section~\ref{part:datasets} provides $23$ classification tasks.
For each task and each model, a separate probing classifier is trained. 
The probing classifier is consistently a random forest with 150 trees. For each latent representation, we report the accuracy achieved by the classifier, using the corresponding audio representation as input. Only accuracy is reported here due to space constraints, as additional metrics (recall, precision, $F_1$) yielded consistent results without altering the conclusions.

As highlighted in Section~\ref{part:datasets}, previous studies have consistently demonstrated the presence of an individual acoustic signature in babies' cries. To mitigate potential biases, such as predicting sex based on individual signatures, we employ a leave-one-baby-out method: the model is trained on all but one baby and tested on the excluded baby. For each baby left out, we report the mean and standard deviation of the accuracies obtained.
For the identity prediction task, we use a $k$-fold cross-validation approach with $k=10$, calculating and reporting the mean and standard deviation across all folds. We apply the same procedure to the Donate A Cry dataset, despite the lack of information on the child’s identity. This limitation will be revisited in the discussion of results.
To address class imbalance in certain datasets, we oversampled the minority classes by generating synthetic observations as needed \cite{chawla_smote_2002}.


\section{Results}\label{part:results}

\begin{figure}[ht!]
\centering
  \subfloat[EnesBabyCries2\label{fig:enes}]{\includegraphics[width=0.5\linewidth]{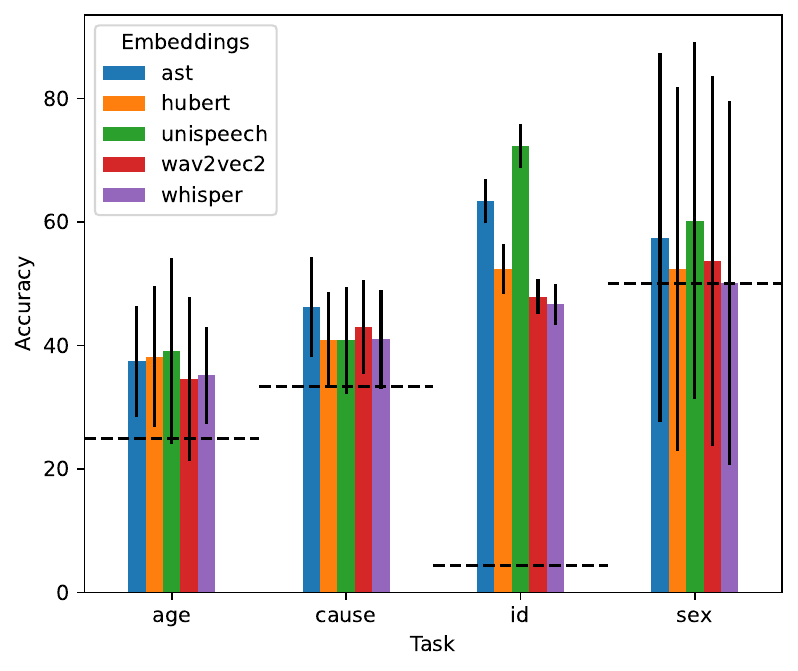}}
  \hfill
  \subfloat[Koutseff\label{fig:koutseff}]{\includegraphics[width=0.5\linewidth]{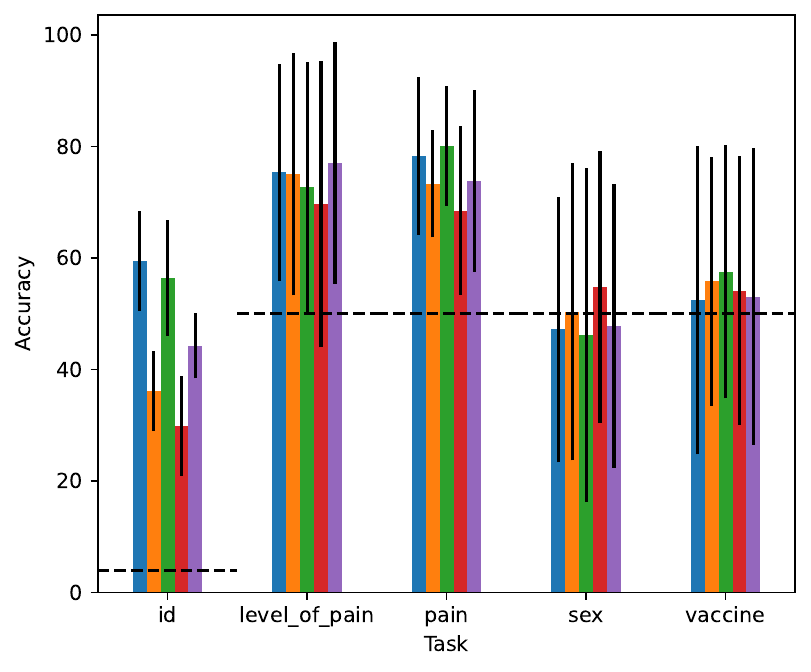}}
  \\
  \subfloat[Cornec\label{fig:cornec}]{\includegraphics[width=0.5\linewidth]{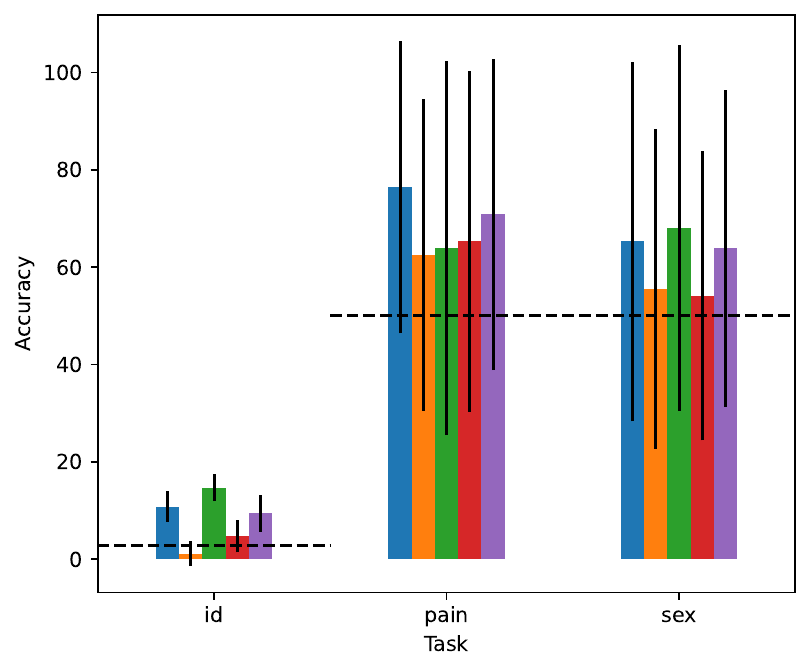}}
  \subfloat[Bouchet\label{fig:bouchet}]{\includegraphics[width=0.5\linewidth]{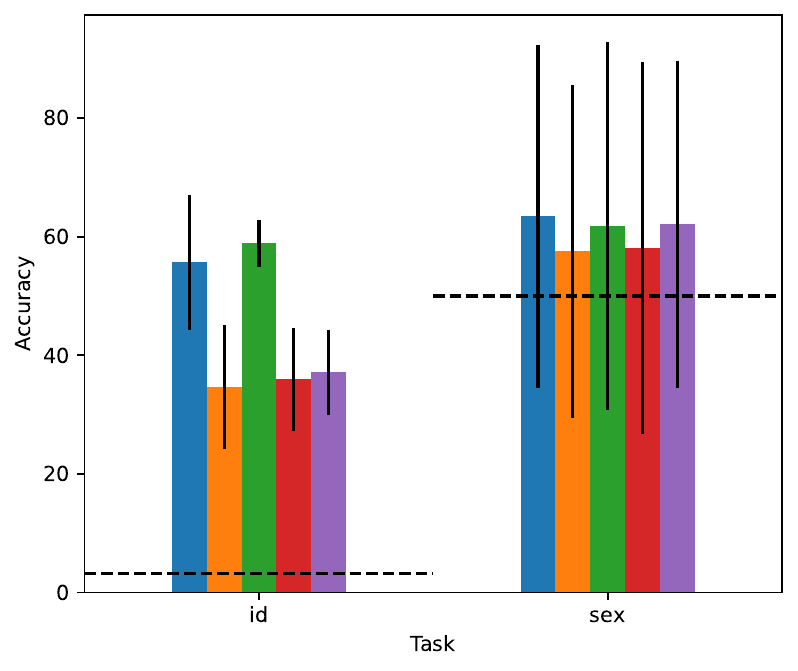}}
  \\
  \subfloat[Lekfir\label{fig:lekfir}]{\includegraphics[width=0.5\linewidth]{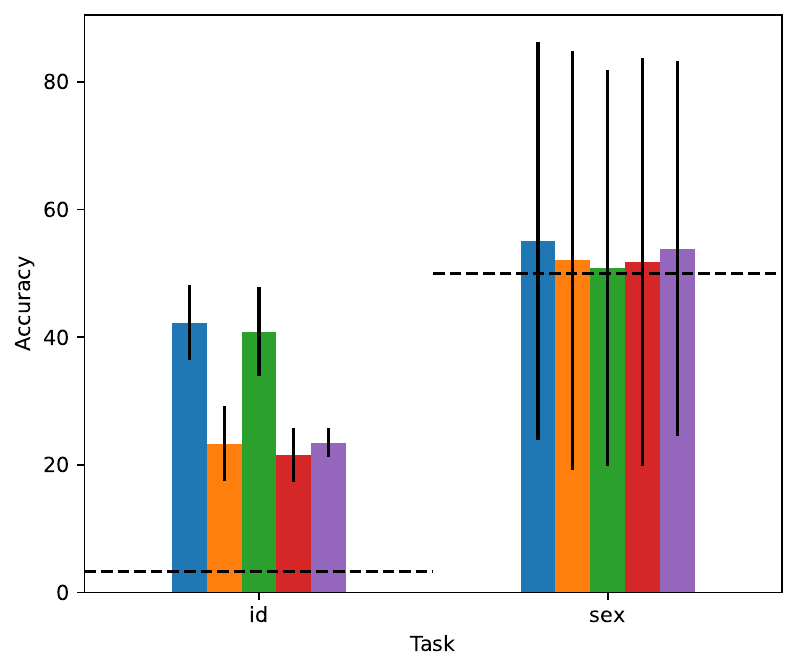}}
  \subfloat[Vial\label{fig:vial}]{\includegraphics[width=0.5\linewidth]{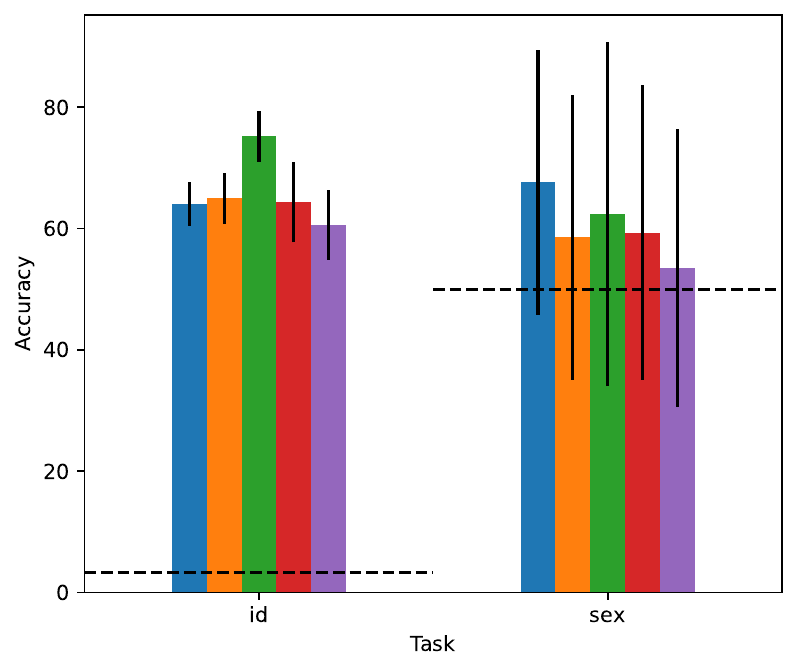}}
  \\
  \subfloat[Donate A Cry\label{fig:dac}]{\includegraphics[width=0.5\linewidth]{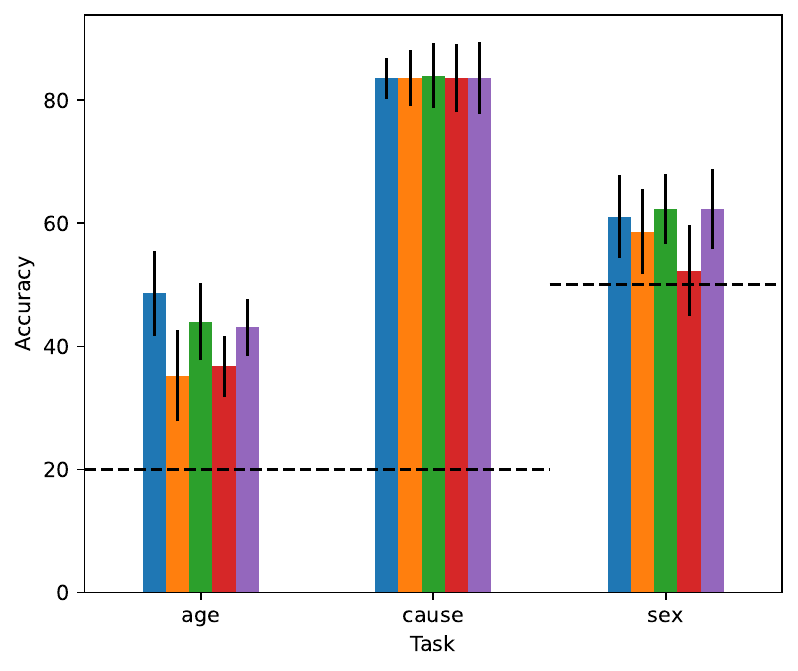}}
  \subfloat[Cry Celeb\label{fig:cry_celeb}]{\includegraphics[width=0.5\linewidth]{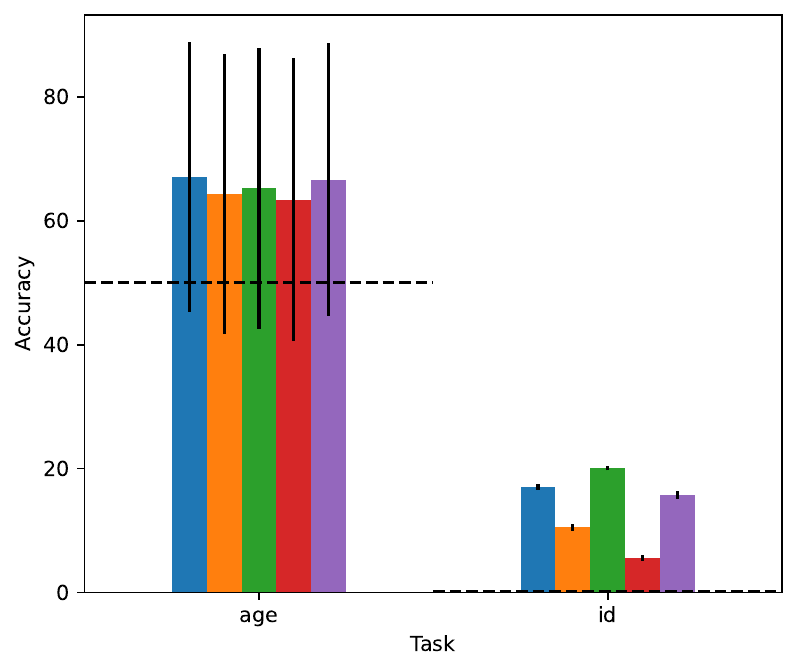}}
  \\
  \caption{Classification accuracy for each dataset and for each embeddings (error bars = standard deviation; horizontal line = chance}
  \label{fig:accuracy} 
\end{figure}

The results are presented in Table~\ref{tab:accuracy} and illustrated in Figure~\ref{fig:accuracy}. Among the twenty-three tasks tested, the mean accuracy and standard deviation exceeded chance for fourteen tasks. As anticipated, sex classification and vaccine type prediction could not be distinguished, accounting for the nine tasks with no significant results.
Of the fourteen tasks where performance surpassed chance, Wav2Vec2 and Unispeech achieved significant results in 78.57\% of cases, HuBERT and Whisper in 92.86\%, and AST consistently outperformed chance. Regarding best-performing models, Unispeech achieved the highest accuracy on seven tasks, AST on four, HuBERT on two, Whisper on one, while Wav2Vec2 did not achieve the best result on any task.

\begin{table*}
\fontsize{8pt}{8pt}\selectfont

\caption{Accuracy for each classification tasks of each dataset, for each model embedding. The classifier is always a random forest with 150 trees. We put in bold when the difference between mean accuracy and standard deviation is strictly higher than chance. We underline, for each task, the embedding that achieves the highest accuracy.}
\label{tab:accuracy}
\setlength\tabcolsep{0pt} 

\begin{tabular*}{\linewidth}{@{\extracolsep{\fill}} cc ccccc}
\toprule
     \multirow{2}{*}{Dataset} & \multirow{2}{*}{Task} & \multicolumn{5}{c}{Accuracy} \\ 
\cmidrule{3-7}
     & & Wav2Vec2 & HuBERT & Whisper & Unispeech & AST \\
\midrule
     \multirow{4}{*}{EnesBabyCries2} & id & $\bm{48 \pm 3}$ & $\bm{52  \pm 4}$ & $\bm{47 \pm 3}$ & \underline{$\bm{72 \pm 4}$} & $\bm{63 \pm 4}$ \\
     & sex & $54 \pm 30$ & $52 \pm 29$ & $50 \pm 29$ & \underline{$60 \pm 29$} & $57 \pm 30$ \\
     & age & $35 \pm 13$ & \underline{$\bm{38 \pm 11}$} & $\bm{35 \pm 8}$ & $39 \pm 15$ & $\bm{37 \pm 9}$\\
     & cause & $\bm{43 \pm 8}$ & $41 \pm 8$ & $41 \pm 8$ & $41 \pm 9$ & \underline{$\bm{46 \pm 8}$} \\
\addlinespace
     \multirow{5}{*}{Koutseff} & id & $\bm{30 \pm 9}$ & $\bm{36 \pm 7}$ & $\bm{44 \pm 6}$ & $\bm{56 \pm 10}$ & \underline{$\bm{59 \pm 9}$}  \\
     & sex & \underline{$55 \pm 24$} & $50 \pm 27$ & $48 \pm 25$ & $46 \pm 30$ & $47 \pm 24$ \\ 
     & pain & $\bm{68 \pm 15}$ & $\bm{73 \pm 10}$ & $\bm{74 \pm 16}$ & \underline{$\bm{80 \pm 11}$} & $\bm{78 \pm 14}$ \\
     & level of pain & $70 \pm 26$ & $\bm{75 \pm 22}$ & \underline{$\bm{77 \pm 22}$} & $73 \pm 23$ & $\bm{75 \pm 20}$  \\
     & vaccine & $54 \pm 24$ & $56 \pm 22$ & $53 \pm 27$ & \underline{$58 \pm 23$} & $52 \pm 28$ \\ 
\addlinespace
    \multirow{3}{*}{Cornec} & id & $\bm{5 \pm 3}$ & $\bm{1 \pm 3}$ & $\bm{9 \pm 4}$ & \underline{$\bm{15 \pm 3}$} & $\bm{11 \pm 3}$ \\
    & sex & $54 \pm 30$ & $56 \pm 33$ & $64 \pm 33$ & \underline{$68 \pm 38$} & $65 \pm 37$ \\
    & pain & $65 \pm 35$ & $62 \pm 32$ & $71 \pm 32$ & $64 \pm 38$ & \underline{$76 \pm 30$} \\
\addlinespace
    \multirow{2}{*}{Bouchet} & id & $\bm{36 \pm 9}$ & $\bm{35 \pm 10}$ & $\bm{37 \pm 7}$ & \underline{$\bm{59 \pm 4}$} & $\bm{56 \pm 11}$  \\
    & sex & $58 \pm 31$ & $58 \pm 28$ & $62 \pm 28$ & $62 \pm 31$ & \underline{$63 \pm 29$} \\ 
\addlinespace
    \multirow{2}{*}{Lefkir} & id & $\bm{22 \pm 4}$ & $\bm{23 \pm 6}$ & $\bm{23 \pm 2}$ & $\bm{41 \pm 7}$ & \underline{$\bm{42 \pm 6}$} \\
    & sex & $52 \pm 32$ & $52 \pm 33$ & $54 \pm 29$ & $51 \pm 31$ & \underline{$55 \pm 31$} \\
\addlinespace
    \multirow{2}{*}{Vial} & id & $\bm{64 \pm 7}$ & $\bm{65 \pm 4}$ & $\bm{61 \pm 6}$ & \underline{$\bm{75 \pm 4}$} & $\bm{64 \pm 4}$  \\
    & sex & $59 \pm 24$ & $59 \pm 23$ & $54 \pm 23$ & $62 \pm 28$ & \underline{$68 \pm 22$} \\
\addlinespace
    \multirow{3}{*}{Donate A Cry} & sex & $52 \pm 7$ & $\bm{59 \pm 7}$ & $\bm{62 \pm 7}$ & \underline{$\bm{62 \pm 6}$} & $\bm{61 \pm 7}$ \\
    & age & $\bm{37 \pm 5}$ & $\bm{35 \pm 7}$ & $\bm{43 \pm 5}$ & $\bm{44 \pm 6}$ & \underline{$\bm{49 \pm 7}$}  \\
    & cause & $\bm{84 \pm 6}$ & \underline{$\bm{86 \pm 5}$} & $\bm{84 \pm 6}$ & $\bm{84 \pm 5}$ & $\bm{84 \pm 3}$ \\
\addlinespace
    \multirow{2}{*}{CryCeleb} & id & $\bm{6 \pm 0}$ & $\bm{11 \pm 1}$ & $\bm{16 \pm 1}$ & \underline{$\bm{20 \pm 0}$} & $\bm{17 \pm 0}$ \\
    & age & $63 \pm 23$ & $64 \pm 23$ & $67 \pm 22$ & $65 \pm 23$ & \underline{$67 \pm 22$} \\
\bottomrule
\end{tabular*}

\end{table*}

\subsection{Identity in cries is encoded by speech models' latent representation}

Identity emerges as the most robustly encoded information, with all models consistently outperforming chance on all identity prediction tasks. While all models encode information related to the individual signature present in cries, Unispeech’s latent representations appear to excel in this domain: of the seven tasks where Unispeech achieved the best results, five were identity prediction tasks. Since the identity of a cry reflects a combination of several acoustic parameters, the speech models successfully captured these parameters in their latent representations.

\subsection{Tonality and melodicity}

The representations do not allow us to perform better than chance on the CryCeleb dataset. However, the difference in this dataset is minimal \textemdash only a few days between birth and discharge from the maternity hospital. If we examine the other two datasets where we test for age, the results demonstrate that the representations also encode age effectively, outperforming chance in 80\% of the cases. While Wav2Vec2 and Unispeech surpass chance in one out of two tasks, HuBERT, Whisper and AST consistently outperform chance. As children’s cries become more tonal and melodic with age, the representations successfully encode these parameters.

\subsection{Non-linear phenomena and roughness}

We evaluated pain encoding using two datasets. None of representations succeeded on the Cornec dataset, but we have just two cries per baby in that dataset. On the other hand, all latent representations performed successfully on the Koutseff dataset, with Unispeech achieving the best result. Moreover, the latent representations of HuBERT, Whisper, and AST effectively encoded pain levels. These representations likely capture information about the non-linear phenomena and the roughness of vocalizations, as pain in cries is closely linked to these acoustic parameters.

\subsection{Cause of the cry}

As an unexpected result, we were able to predict the cause of cries with accuracy better than chance. As noted in Section~\ref{part:datasets}, it is challenging to draw definitive conclusions from the results on the Donate A Cry dataset, as the lack of information about the babies’ identities introduces a huge bias. On the other hand, the representations from Wav2Vec2 and AST were able to distinguish cries caused by hunger, discomfort, and loneliness beyond chance on the EnesBabyCries2 dataset. This suggests that these representations may capture features that previous investigations have thus far been unable to identify \cite{lockhart-bouron_infant_2023}.

\section{Discussion}\label{part:discussion}

\subsection{Speech models' latent representation and source-filter theory}

As expected, the latent representations do not distinguish the child’s sex. This aligns with the lack of acoustic sex differences in cries \cite{reby_sex_2016, cornec_human_2024}, consistent with the absence of sexual dimorphism before puberty \cite{barbier_human_2015}. The representations successfully differentiate pain cries, which are characterized by non-linear phenomena and are rougher than non-pain cries \cite{koutseff_acoustic_2018, cornec_human_2024, corvin_pain_2024}. These phenomena indicate increased tension in the vocal signal source, reflecting chaotic vibrations of the vocal folds \cite{fitch_calls_2002, anikin_nlp}. This suggests that latent representations capture information about the tension in the vocal signal source, which could be valuable for other tasks, such as emotion recognition.

\subsection{Individual signature in cry and speech}

Our results demonstrate that latent representations learned in a self-supervised manner on speech data contain information for detecting the individual signature present in babies' cries \textemdash a signature which remains stable throughout development \cite{lockhart-bouron_infant_2023}. These representations are also known to encode information useful for speaker recognition tasks \cite{novoselov_robust_2022}. This concordance suggests a continuity between speech and cries in terms of individual signature: the acoustic parameters that signal individuality in speech might be similar as those in cries. Unispeech’s superior performance on this task supports this hypothesis. The model learns latent representations in a self-supervised manner while a supervised task explicitly aligns the learned representations with known phonetic structures \cite{wang_unispeech_2021}. This suggests that the individual signature begins to form from birth.

\subsection{Transferability and training}

Taken together, the results highlight the value of transferring speech model representations to the study of baby cries, while also revealing differences between the representations. Whisper, trained on the largest dataset among the compared models but in a supervised manner, does not achieve the best results. Its representations are tailored for Indo-European languages, with reduced transferability to tasks involving languages that differ significantly from its training data \cite{radford_robust_2023}. Supervised training encourages it to learn representations closely aligned with speech-specific features, whereas the self-supervised training of other models enables them to learn more generalized, ``looser” representations, which are more transferable to non-speech domains, such as cries.
AST, on the other hand, which was also trained in a supervised manner \cite{gong_ast_2021}, consistently outperforms chance across all tasks. However, its training set, which includes AudioSet, already contains cry data \cite{gemmeke_audio_2017}, likely contributing to its success. For future research, it would be interesting to fine-tune the self-supervised speech models, capable of learning representations useful for cry classification from speech alone, on datasets specifically focused on cries.

\subsection{Limits}

Our results do not clarify how the latent representations of these models encode information useful for determining the identity of the baby or the roughness of the cry. Additionally, while we provide insights into modeling choices aimed at learning more universal representations, we cannot demonstrate a clear superiority of one model over another. Further analysis is necessary to address these questions.

Although our results might suggest that these latent representations could be used to predict the cause of cries, we must remain very cautious with this possibility. The results obtained from the Donate A Cry dataset are unreliable, particularly due to the lack of information about the identity of the crying child. 
The identity of the baby is well established in the EnesBabyCries2 dataset.
It is possible that the latent representations used here may capture information that the approach used in the initial study \cite{lockhart-bouron_infant_2023} did not allow to access to. It is also likely that the levels of distress encoded by the three causes labeled in the initial study (discomfort, hunger, and isolation) may vary somewhat between causes. Our classification, rather than properly identifying the causes, would then simply reflect these tendencies to correspond to varying levels of discomfort or pain. 
We therefore remain unconvinced, at this stage, that the cause of a cry can be identified by our approach. 

\section{Conclusion}\label{part:conclusion}

In conclusion, we have shown that latent representations of speech models contain valuable information for the classification of human infant cries. They reliably encode identity and age of the crying baby. In addition, their application to cry classification revealed that these latent representations also capture information about the tension in the vocal signal source. These results encourage the broader use of these models to study baby cries.

\bibliographystyle{IEEEtran}
\bibliography{transformers}

\begin{thebibliography}{10}
\providecommand{\url}[1]{#1}
\csname url@samestyle\endcsname
\providecommand{\newblock}{\relax}
\providecommand{\bibinfo}[2]{#2}
\providecommand{\BIBentrySTDinterwordspacing}{\spaceskip=0pt\relax}
\providecommand{\BIBentryALTinterwordstretchfactor}{4}
\providecommand{\BIBentryALTinterwordspacing}{\spaceskip=\fontdimen2\font plus
\BIBentryALTinterwordstretchfactor\fontdimen3\font minus \fontdimen4\font\relax}
\providecommand{\BIBforeignlanguage}[2]{{%
\expandafter\ifx\csname l@#1\endcsname\relax
\typeout{** WARNING: IEEEtran.bst: No hyphenation pattern has been}%
\typeout{** loaded for the language `#1'. Using the pattern for}%
\typeout{** the default language instead.}%
\else
\language=\csname l@#1\endcsname
\fi
#2}}
\providecommand{\BIBdecl}{\relax}
\BIBdecl

\bibitem{gustafsson_fathers_2013}
\BIBentryALTinterwordspacing
E.~Gustafsson, F.~Levréro, D.~Reby, and N.~Mathevon, ``\BIBforeignlanguage{en}{Fathers are just as good as mothers at recognizing the cries of their baby},'' \emph{\BIBforeignlanguage{en}{Nature Communications}}, vol.~4, no.~1, p. 1698, Apr. 2013, publisher: Nature Publishing Group. [Online]. Available: \url{https://www.nature.com/articles/ncomms2713}
\BIBentrySTDinterwordspacing

\bibitem{bouchet_baby_2020}
\BIBentryALTinterwordspacing
H.~Bouchet, A.~Plat, F.~Levréro, D.~Reby, H.~Patural, and N.~Mathevon, ``\BIBforeignlanguage{en}{Baby cry recognition is independent of motherhood but improved by experience and exposure},'' \emph{\BIBforeignlanguage{en}{Proceedings of the Royal Society B: Biological Sciences}}, vol. 287, no. 1921, p. 20192499, Feb. 2020. [Online]. Available: \url{https://royalsocietypublishing.org/doi/10.1098/rspb.2019.2499}
\BIBentrySTDinterwordspacing

\bibitem{lockhart-bouron_infant_2023}
\BIBentryALTinterwordspacing
M.~Lockhart-Bouron, A.~Anikin, K.~Pisanski, S.~Corvin, C.~Cornec, L.~Papet, F.~Levréro, C.~Fauchon, H.~Patural, D.~Reby, and N.~Mathevon, ``\BIBforeignlanguage{en}{Infant cries convey both stable and dynamic information about age and identity},'' \emph{\BIBforeignlanguage{en}{Communications Psychology}}, vol.~1, no.~1, pp. 1--15, Oct. 2023, number: 1 Publisher: Nature Publishing Group. [Online]. Available: \url{https://www.nature.com/articles/s44271-023-00022-z}
\BIBentrySTDinterwordspacing

\bibitem{koutseff_acoustic_2018}
\BIBentryALTinterwordspacing
A.~Koutseff, D.~Reby, O.~Martin, F.~Levrero, H.~Patural, and N.~Mathevon, ``\BIBforeignlanguage{en}{The acoustic space of pain: cries as indicators of distress recovering dynamics in pre-verbal infants},'' \emph{\BIBforeignlanguage{en}{Bioacoustics}}, vol.~27, no.~4, pp. 313--325, Oct. 2018. [Online]. Available: \url{https://www.tandfonline.com/doi/full/10.1080/09524622.2017.1344931}
\BIBentrySTDinterwordspacing

\bibitem{orlandi_automatic_2012}
S.~Orlandi, C.~Manfredi, L.~Bocchi, and M.~L. Scattoni, ``\BIBforeignlanguage{eng}{Automatic newborn cry analysis: a non-invasive tool to help autism early diagnosis},'' \emph{\BIBforeignlanguage{eng}{Annual International Conference of the IEEE Engineering in Medicine and Biology Society. IEEE Engineering in Medicine and Biology Society. Annual International Conference}}, vol. 2012, pp. 2953--2956, 2012.

\bibitem{esposito_cry_2017}
G.~Esposito, N.~Hiroi, and M.~L. Scattoni, ``Cry, baby, cry: {Expression} of distress as a biomarker and modulator in autism spectrum disorder,'' \emph{International Journal of Neuropsychopharmacology}, vol.~20, no.~6, pp. 498--503, 2017, place: United Kingdom Publisher: Oxford University Press.

\bibitem{khozaei_early_2020}
\BIBentryALTinterwordspacing
A.~Khozaei, H.~Moradi, R.~Hosseini, H.~Pouretemad, and B.~Eskandari, ``\BIBforeignlanguage{en}{Early screening of autism spectrum disorder using cry features},'' \emph{\BIBforeignlanguage{en}{PLOS ONE}}, vol.~15, no.~12, p. e0241690, Dec. 2020, publisher: Public Library of Science. [Online]. Available: \url{https://journals.plos.org/plosone/article?id=10.1371/journal.pone.0241690}
\BIBentrySTDinterwordspacing

\bibitem{lawford_acoustic_2022}
H.~L.~S. Lawford, H.~Sazon, C.~Richard, M.~P. Robb, and S.~Bora, ``\BIBforeignlanguage{eng}{Acoustic {Cry} {Characteristics} of {Infants} as a {Marker} of {Neurological} {Dysfunction}: {A} {Systematic} {Review} and {Meta}-{Analysis}},'' \emph{\BIBforeignlanguage{eng}{Pediatric Neurology}}, vol. 129, pp. 72--79, Apr. 2022.

\bibitem{mohamed_self-supervised_2022}
\BIBentryALTinterwordspacing
A.~Mohamed, H.-y. Lee, L.~Borgholt, J.~D. Havtorn, J.~Edin, C.~Igel, K.~Kirchhoff, S.-W. Li, K.~Livescu, L.~Maaløe, T.~N. Sainath, and S.~Watanabe, ``Self-{Supervised} {Speech} {Representation} {Learning}: {A} {Review},'' \emph{IEEE Journal of Selected Topics in Signal Processing}, vol.~16, no.~6, pp. 1179--1210, Oct. 2022, conference Name: IEEE Journal of Selected Topics in Signal Processing. [Online]. Available: \url{https://ieeexplore.ieee.org/document/9893562}
\BIBentrySTDinterwordspacing

\bibitem{vaswani_attention_2017}
\BIBentryALTinterwordspacing
A.~Vaswani, N.~Shazeer, N.~Parmar, J.~Uszkoreit, L.~Jones, A.~N. Gomez, L.~Kaiser, and I.~Polosukhin, ``Attention is {All} you {Need},'' in \emph{Advances in {Neural} {Information} {Processing} {Systems}}, vol.~30.\hskip 1em plus 0.5em minus 0.4em\relax Curran Associates, Inc., 2017. [Online]. Available: \url{https://papers.nips.cc/paper_files/paper/2017/hash/3f5ee243547dee91fbd053c1c4a845aa-Abstract.html}
\BIBentrySTDinterwordspacing

\bibitem{cauzinille_investigating_2024}
\BIBentryALTinterwordspacing
J.~Cauzinille, B.~Favre, R.~Marxer, D.~Clink, A.~H. Ahmad, and A.~Rey, ``Investigating self-supervised speech models' ability to classify animal vocalizations: {The} case of gibbon's vocal signatures,'' in \emph{Interspeech 2024}, 2024, pp. 132--136. [Online]. Available: \url{https://www.isca-archive.org/interspeech_2024/cauzinille24_interspeech.html}
\BIBentrySTDinterwordspacing

\bibitem{cornec_human_2024}
\BIBentryALTinterwordspacing
C.~Cornec, N.~Mathevon, K.~Pisanski, D.~Entani, C.~Monghiemo, B.~Bola, V.~Planas-Bielsa, D.~Reby, and F.~Levréro, ``\BIBforeignlanguage{en}{Human infant cries communicate distress and elicit sex stereotypes: {Cross} cultural evidence},'' \emph{\BIBforeignlanguage{en}{Evolution and Human Behavior}}, vol.~45, no.~1, pp. 48--57, Jan. 2024. [Online]. Available: \url{https://linkinghub.elsevier.com/retrieve/pii/S1090513823000715}
\BIBentrySTDinterwordspacing

\bibitem{corvin_pain_2024}
\BIBentryALTinterwordspacing
S.~Corvin, C.~Fauchon, H.~Patural, R.~Peyron, D.~Reby, F.~Theunissen, and N.~Mathevon, ``\BIBforeignlanguage{English}{Pain cues override identity cues in baby cries},'' \emph{\BIBforeignlanguage{English}{iScience}}, vol.~0, no.~0, Jun. 2024, publisher: Elsevier. [Online]. Available: \url{https://www.cell.com/iscience/abstract/S2589-0042(24)01600-6}
\BIBentrySTDinterwordspacing

\bibitem{anikin_nlp}
A.~Anikin, D.~Reby, and K.~Pisanski, ``Nonlinear vocal phenomena and speech intelligibility,'' \emph{Philosophical Transactions B}, (in press).

\bibitem{budaghyan_cryceleb_2024}
\BIBentryALTinterwordspacing
D.~Budaghyan, C.~C. Onu, A.~Gorin, C.~Subakan, and D.~Precup, ``{CryCeleb}: {A} {Speaker} {Verification} {Dataset} {Based} on {Infant} {Cry} {Sounds},'' in \emph{{ICASSP} 2024 - 2024 {IEEE} {International} {Conference} on {Acoustics}, {Speech} and {Signal} {Processing} ({ICASSP})}.\hskip 1em plus 0.5em minus 0.4em\relax Seoul, Korea, Republic of: IEEE, Apr. 2024, pp. 11\,966--11\,970. [Online]. Available: \url{https://ieeexplore.ieee.org/document/10448292/}
\BIBentrySTDinterwordspacing

\bibitem{baevski_wav2vec_2020}
\BIBentryALTinterwordspacing
A.~Baevski, Y.~Zhou, A.~Mohamed, and M.~Auli, ``wav2vec 2.0: {A} {Framework} for {Self}-{Supervised} {Learning} of {Speech} {Representations},'' in \emph{Advances in {Neural} {Information} {Processing} {Systems}}, vol.~33.\hskip 1em plus 0.5em minus 0.4em\relax Curran Associates, Inc., 2020, pp. 12\,449--12\,460. [Online]. Available: \url{https://proceedings.neurips.cc/paper/2020/hash/92d1e1eb1cd6f9fba3227870bb6d7f07-Abstract.html}
\BIBentrySTDinterwordspacing

\bibitem{hsu_hubert_2021}
\BIBentryALTinterwordspacing
W.-N. Hsu, B.~Bolte, Y.-H.~H. Tsai, K.~Lakhotia, R.~Salakhutdinov, and A.~Mohamed, ``{HuBERT}: {Self}-{Supervised} {Speech} {Representation} {Learning} by {Masked} {Prediction} of {Hidden} {Units},'' \emph{IEEE/ACM Transactions on Audio, Speech, and Language Processing}, vol.~29, pp. 3451--3460, 2021, conference Name: IEEE/ACM Transactions on Audio, Speech, and Language Processing. [Online]. Available: \url{https://ieeexplore.ieee.org/document/9585401}
\BIBentrySTDinterwordspacing

\bibitem{radford_robust_2023}
A.~Radford, J.~W. Kim, T.~Xu, G.~Brockman, C.~McLeavey, and I.~Sutskever, ``Robust speech recognition via large-scale weak supervision,'' in \emph{Proceedings of the 40th {International} {Conference} on {Machine} {Learning}}, ser. {ICML}'23, vol. 202.\hskip 1em plus 0.5em minus 0.4em\relax Honolulu, Hawaii, USA: JMLR.org, Jul. 2023, pp. 28\,492--28\,518.

\bibitem{wang_unispeech_2021}
\BIBentryALTinterwordspacing
C.~Wang, Y.~Wu, Y.~Qian, K.~Kumatani, S.~Liu, F.~Wei, M.~Zeng, and X.~Huang, ``\BIBforeignlanguage{en}{{UniSpeech}: {Unified} {Speech} {Representation} {Learning} with {Labeled} and {Unlabeled} {Data}},'' in \emph{\BIBforeignlanguage{en}{Proceedings of the 38th {International} {Conference} on {Machine} {Learning}}}.\hskip 1em plus 0.5em minus 0.4em\relax PMLR, Jul. 2021, pp. 10\,937--10\,947, iSSN: 2640-3498. [Online]. Available: \url{https://proceedings.mlr.press/v139/wang21y.html}
\BIBentrySTDinterwordspacing

\bibitem{gong_ast_2021}
\BIBentryALTinterwordspacing
Y.~Gong, Y.-A. Chung, and J.~Glass, ``{AST}: {Audio} {Spectrogram} {Transformer},'' in \emph{Interspeech 2021}, 2021, pp. 571--575. [Online]. Available: \url{https://www.isca-archive.org/interspeech_2021/gong21b_interspeech.html}
\BIBentrySTDinterwordspacing

\bibitem{chawla_smote_2002}
\BIBentryALTinterwordspacing
N.~V. Chawla, K.~W. Bowyer, L.~O. Hall, and W.~P. Kegelmeyer, ``\BIBforeignlanguage{en}{{SMOTE}: {Synthetic} {Minority} {Over}-sampling {Technique}},'' \emph{\BIBforeignlanguage{en}{Journal of Artificial Intelligence Research}}, vol.~16, pp. 321--357, Jun. 2002. [Online]. Available: \url{https://www.jair.org/index.php/jair/article/view/10302}
\BIBentrySTDinterwordspacing

\bibitem{reby_sex_2016}
\BIBentryALTinterwordspacing
D.~Reby, F.~Levréro, E.~Gustafsson, and N.~Mathevon, ``Sex stereotypes influence adults’ perception of babies’ cries,'' \emph{BMC Psychology}, vol.~4, no.~1, p.~19, Apr. 2016. [Online]. Available: \url{https://doi.org/10.1186/s40359-016-0123-6}
\BIBentrySTDinterwordspacing

\bibitem{barbier_human_2015}
\BIBentryALTinterwordspacing
G.~Barbier, L.-J. Boë, G.~Captier, and R.~Laboissière, ``\BIBforeignlanguage{en}{Human vocal tract growth: {A} longitudinal study of the development of various anatomical structures},'' Sep. 2015. [Online]. Available: \url{https://hal.science/hal-01200990}
\BIBentrySTDinterwordspacing

\bibitem{fitch_calls_2002}
\BIBentryALTinterwordspacing
W.~T. Fitch, J.~Neubauer, and H.~Herzel, ``Calls out of chaos: the adaptive significance of nonlinear phenomena in mammalian vocal production,'' \emph{Animal Behaviour}, vol.~63, no.~3, pp. 407--418, Mar. 2002. [Online]. Available: \url{https://www.sciencedirect.com/science/article/pii/S0003347201919128}
\BIBentrySTDinterwordspacing

\bibitem{novoselov_robust_2022}
\BIBentryALTinterwordspacing
S.~Novoselov, G.~Lavrentyeva, A.~Avdeeva, V.~Volokhov, and A.~Gusev, ``\BIBforeignlanguage{en}{Robust {Speaker} {Recognition} with {Transformers} {Using} wav2vec 2.0},'' Mar. 2022, arXiv:2203.15095 [cs]. [Online]. Available: \url{http://arxiv.org/abs/2203.15095}
\BIBentrySTDinterwordspacing

\bibitem{gemmeke_audio_2017}
J.~F. Gemmeke, D.~P.~W. Ellis, D.~Freedman, A.~Jansen, W.~Lawrence, R.~C. Moore, M.~Plakal, and M.~Ritter, ``Audio {Set}: {An} ontology and human-labeled dataset for audio events,'' in \emph{2017 {IEEE} {International} {Conference} on {Acoustics}, {Speech} and {Signal} {Processing} ({ICASSP})}, Mar. 2017, pp. 776--780, iSSN: 2379-190X.

\end{thebibliography}

\end{document}